\documentclass[aps,pra,reprint,superscriptaddress,showpacs,longbibliography]{revtex4-1}

\usepackage{graphicx}
\usepackage{printlen}
\usepackage{color}
\usepackage{amsmath,amssymb}
\usepackage{bm}
\usepackage{upgreek}
\usepackage{xspace}
\usepackage[colorlinks,urlcolor=blue,citecolor=blue,linkcolor=blue]{hyperref}
\usepackage[normalem]{ulem}
\graphicspath{{figures/}}
\usepackage{enumerate}
\usepackage{gensymb}

\newcommand{\Rb}{$^{87}$Rb\xspace}

\newcommand{\reffig}[1]{\mbox{Fig.~\ref{fig:#1}}}
\newcommand{\refeq}[1]{\mbox{Eq.~\ref{eq:#1}}}

\newcommand{\ket}[1]{\left| #1 \right \rangle \xspace}

\newcommand{\unit}[1]{\,\mathrm{#1}}

\begin{document}

\title{Readout delay free Bragg atom interferometry using overlapped spatial fringes}

\author{P.~B.~Wigley}
\author{K.~S.~Hardman}
\author{C.~Freier}
\author{P.~J.~Everitt}
\author{S.~Legge}
\author{P.~Manju}
\author{J.~D.~Close}
\author{N.~P.~Robins}
\affiliation{Department of Quantum Science, The Australian National University, ACT 0200 Australia}

\date{\today}

\begin{abstract}
A method for mitigating the readout delay characteristic of Bragg-based atom interferometry is presented, utilizing an asymmetric Mach-Zehnder interferometer sequence to generate spatial fringes that are read out while still overlapped. The time of flight after the final beamsplitter is engineered to facilitate constructive overlap of the fringes on the output states. A noise analysis performed using a precision atom interferometer is presented, comparing the traditional separated symmetric scheme with that of the separated and overlapped asymmetric scheme showing no significant increase in short term phase noise.
\end{abstract}

% 03.75.Lm: Solitons in Bose-Einstein condensates
% 03.75.Mn: Bose-Einstein condensation, multicomponent and spinor condensates 
% 03.75.-b: Matter waves
\maketitle

% \section{Introduction}

	Since its inception, atom interferometry has proved an invaluable tool, enabling some of the most accurate measurements of physical quantities to date including local gravity \cite{freier_mobile_2016,menoret_gravity_2018}, the gravitational constant \cite{rosi_precision_2014}, the fine structure constant \cite{parker_measurement_2018}, and the ratio of Planck's constant to the atomic mass \cite{weiss_precision_1993,bouchendira_new_2011}.
	% With their increasing sensitivity, atom interferometers have been proposed as detectors of gravitational waves \cite{dimopoulos_atomic_2008,chaibi_low_2016}.  
	The high precision and high accuracy of atom interferometers coupled with their inherent long-term stability and calibration have seen them become increasingly competitive with traditional sensors used in mineral exploration and navigation \cite{stockton_absolute_2011,canuel_six-axis_2006,bidel_compact_2013}. To this end, much progress has been made toward improving size \cite{bodart_cold_2010,bidel_compact_2013}, bandwidth \cite{mcguinness_high_2012} as well as field readiness \cite{stern_light-pulse_2009} including through the hybridization of atom interferometers with classical sensors \cite{lautier_hybridizing_2014,cheiney_navigation-compatible_2018}. 

	\begin{figure}
		\centering\includegraphics[width=\columnwidth]{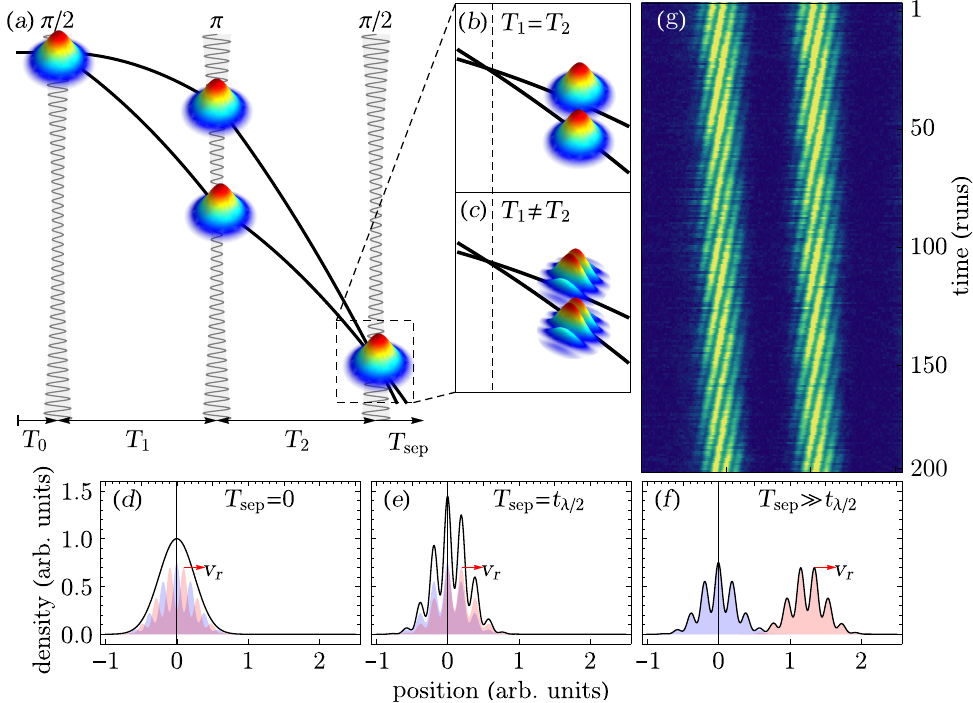}
		\caption{(a) Diagram of a Mach-Zehnder atom-interferometer sequence. An atom source is prepared and released into freefall. A Bragg lattice is used to split this into two momentum states at $T_0$. After time $T_1$ a mirror pulse is applied swapping the momentum states. After a subequent time $T_2$, a second beamsplitter pulse is applied to recombine and interfere the atoms. Different output states are generated depending on $\delta T = T_1 - T_2$. (b) When $\delta T=0$ the momentum states are output with the relative population providing phase information. (c) If $\delta T\neq 0$ a spatial interference pattern is observed with the phase of the interference providing information on the difference in paths. Parts (d), (e) and (f) illustrate the operation of the overlapped and separated spatial fringe methods. At the final beamsplitter (d), the two states are overlapped and out of phase, no spatial modulation oberseved. Waiting $t_{\pi/2}$ (e) results in the two states moving spatially into phase with the resulting image showing fringes with increased signal-to-noise. (f) the two output states after fulling separating, with (g) displaying example output data over $200$ runs of a $T=50\unit{ms}$ interferometer as the laser phase is scanned by $16\unit{degrees}$ each run, with the fringe phase seen to be stable over many iterations of the experiment.}
		\label{fig:diagram}
	\end{figure}

	Most precision atom interferometers employ Raman transitions to generate the mirror and beamsplitters required for the interferometry sequence. This two-photon process couples atoms to different internal atomic states in addition to providing a momentum kick. Whilst different internal atomic states allow the interferometer outputs to be measured independently without requiring state separation, they introduce a susceptibility to systematic shifts from electromagnetic fields and magnetic field gradients and require careful shielding of the apparatus from external fields. In contrast, Bragg transitions couple only between momentum states allowing such precision interferometers to operate unshielded from the environment and enabling simultaneous precision measurements of magnetic field gradient and gravitational acceleration \cite{hardman_simultaneous_2016}. These benefits emphasize the potential of Bragg-based atom interferometers and advance the possibility of an all-in-one atomic sensor. In the transition from lab-based devices to those that are field deployable, size, weight, and power (SWaP) becomes an increasingly important metric with the readout delay inherent to Bragg atom interferometers limiting sensitivity for a given size. Large momentum transfer (LMT) interferometry mitigates this effect somewhat \cite{mcdonald_80hbark_2013,chiow_102hbark_2011}, but also scales the laser phase noise and power requirements \cite{szigeti_why_2012}. This manuscript presents a Bragg-based technique that removes the requirement of separation and therefore operates with no readout delay. The technique is based on the asymmetric Mach-Zehnder interferometer, a scheme that results in a spatial interference pattern that additionally allows for single shot phase readout \cite{muntinga_interferometry_2013,miller_high-contrast_2005,andrews_observation_1997,sugarbaker_enhanced_2013}. By operating the interferometer in a way that allows the the spatial fringes to overlap constructively, the readout delay due to separation is reduced significantly. Coupled with the spatio-temporal coherence provided by a Bose-Einstein condensate (BEC) \cite{hardman_role_2014}, the technique enables precision atom interferometers based on spatial fringes with the possibility of simultaneous measurements of other fields. Although the potential of overlapping spatial fringes has been proposed by Rasel \cite{muntinga_interferometry_2013}, we use a precision device to provide a direct comparison of the overlapped scheme to the symmetric Mach-Zehnder scheme and assess the impact on SWaP. 

% \section{Apparatus}\label{sec:apparatus}

    The apparatus used for this comparison is described in detail in \cite{hardman_simultaneous_2016}. Briefly, a thermal sample of \Rb is confined and cooled using a two-dimensional magneto-optical trap (MOT) before being transferred through a high impedance line to an ultrahigh vacuum 3DMOT where further cooling is achieved. After $6\unit{s}$ a sample of $5\times 10^9$ atoms is acquired before undergoing a compression and polarization gradient cooling stage lowering the temperature to ${\sim20}\unit{\upmu K}$. The atoms are then loaded into a hybrid magnetic quadrupole and crossed optical dipole trap, where an initial stage of evaporation is completed using a microwave knife over $4.5\unit{s}$ leaving $4\times 10^7$\,atoms at $4\unit{\upmu K}$. The magnetic field gradient is subsequently decreased from $150\unit{G/cm}$ to $25\unit{G/cm}$ over $200\unit{ms}$ until the atoms are no longer supported against gravity, allowing efficient loading into the crossed optical dipole trap with the magnetic field subsequently extinguished. The use of this hybrid trap facilitates spatial mode matching from the MOT stage to the dipole trap stage, enabling a higher final atom number to be achieved. The optical trap is generated by a pair of $1064\unit{nm}$ broad linewidth fiber lasers intersecting at $22.5^\circ$ each with waists of $300\unit{\upmu m}$. By reducing the intensity of both beams over $2\unit{s}$, forced evaporation results in a pure $\ket{F=1, m_f=-1}$ $2\times 10^6$\,atom condensate with an in-trap width of approximately $50\unit{\upmu m}$ and an effective temperature of ${\sim50}\unit{nK}$ measured through time-of-flight expansion. The optical trap is then removed and the condensate allowed to fall under gravity. 

    A magnetic superposition state of $\ket{m_f=1,0,-1}$ is generated using Raman transitions through the use of a pair of far-detuned, co-propagating beams with linear and circular polarization, pulsed on $2.2\unit{ms}$ after release from trap. A vertically oriented Bragg lattice is generated with two frequency shifted beams of orthogonal polarization, allowing each magnetic substate to be put into a momentum superposition state, forming the interferometer components. Though all three magnetic substates could form unique simultaneous interferometers, the following study only considers the magnetically insensitive $m_F=0$ state.

    The vertically oriented Mach-Zehnder interferometer (MZI) consists of three ${\sim10\unit{GHz}}$ detuned Bragg pulses, with one beam frequency chirped to account for an increasing Doppler shift resulting from the accelerating atoms. The other beam is adjusted to address the resonance frequency required for transfer of $2\hbar k$ of momentum, where $k=2\pi/\lambda$ is the wavenumber of the light of wavelength $\lambda$ and $\hbar$ is the reduced Planck constant. 
    %The independent frequency control is provided by two acousto-optic modulators driven by a direct digital synthesizer referencing a cesium primary frequency standard. The optical setup at the science head ensures the Bragg transitions are driven by circularly polarized light. The beam is initially aligned to vertical using a liquid mercury mirror and subsequently backcoupled into the fiber. 
    The light itself is generated through a frequency doubled $1560\unit{nm}$ fiber amplifier system using a seed with a narrow $10\unit{Hz}$ linewidth. Bragg pulses with $50\unit{\upmu s}$ full width half maximum are used to couple the cloud to the $2\hbar k$ momentum state. The inertial reference is provided by a custom retro-reflector passively isolated from ground vibrations through the low frequency mechanical oscillator of a geometric anti-spring system \cite{hardman_bec_2016}. Finally, a Stern-Gerlach pulse is used to separate the different magnetic substates at the culmination of the interferometer sequence. In the overlapped asymmetric scheme, this pulse is applied before the interferometer begins, as the noise analysis is only concerned with the magnetically insensitive states. In this case, the other two magnetic substates do not participate in simultaneous interferometers. The Stern-Gerlach pulse is applied outside the interferometer region so as not to introduce phase noise to the interferometer. 

%	    Bragg interferometry \cite{giltner_atom_1995}.

    The apparatus allows for up to ${\sim730\unit{ms}}$ time-of-flight (TOF) with a mix of imaging techniques available for a number of fall times. Absorption imaging on CCD cameras allows for 2D images using interferometer times up to $T=90\unit{ms}$, while frequency modulation imaging (FMI) \cite{hardman_time--flight_2016} allows for 1D density signals for all other possible interferometer times.  

    The Mach-Zehnder atom interferometer has been well studied and described in detail elsewhere \cite{kasevich_measurement_1992}. Briefly, a cold or ultracold atomic source is released from a trapping potential and allowed to freely fall under gravity. A sequence of pulses in a $\pi/2$, $\pi$, $\pi/2$ configuration, separated by $T_1$ and $T_2$, couple vertical momentum states as shown in top of \reffig{diagram}. This results in two output ports, with the total phase difference between the two paths of the interferometer incorporating three main components,
	\begin{align}
		\phi_\text{total} = \phi_\text{propagation} + \phi_\text{laser} + \phi_\text{separation}
	\end{align}
	where  $\phi_\text{propagation}$ corresponds to the phase acquired during the propagation of the states, $\phi_\text{laser}$ corresponds to the phase acquired during the beamsplitter and mirror pulses and $\phi_\text{separation}$ corresponds to a phase due to some finite separation of the output states at the final beamsplitter. In the case of the symmetric Mach-Zehnder interferometer, $T_1=T_2$ and the separation phase component is typically neglected \footnote{Certain cases exist where the separation phase is still generated, such as through phase shear \cite{sugarbaker_enhanced_2013}.}, resulting in a phase of 
    \begin{align}
    	\phi &= \phi_\text{propagation} + \phi_\text{laser}
    		\\&= n \left(\mathbf{k}_\text{eff}\cdot\mathbf{g}-2\pi\alpha\right)T^2+\phi_\text{laser}
    	\label{eq:inertialPhaseSignal}
    \end{align}
    where $n$ is the Bragg order, $T$ is the time between the interferometer pulses (as illustrated in \reffig{diagram}), $k_\text{eff}=4\pi/\lambda$ is the effective wavevector of the optical beamsplitters and mirrors formed using light of wavelength $\lambda$. In order to compensate for the Doppler shift of the falling atoms, the frequency of the laser pulses is swept at a rate $\alpha$, and $\phi_\text{laser}=\phi_1-2\phi_2+\phi_3$ provides the phase of the laser pulses in the lab frame, relative to the inertial reference. The relative population in each of the two output states can be monitored and the phase extracted and related to an inertial acceleration. This requires a delay before measurement to allow the output states to separate. 

	If instead the pulse scheme operates asymmetrically, such that there exists a temporal mismatch, $T_2=T_1\pm\delta T$, the additional separation phase plays a key role. In general the separation phase on each output consists of two components \cite{hogan_light-pulse_2008}, 
	\begin{align}
		\phi_\text{separation} = \frac{\Delta \mathbf{x}}{\hbar}\cdot\mathbf{\bar{p}} + \frac{\Delta \mathbf{p}}{\hbar}\cdot\mathbf{x}
		\label{eq:separationPhase}
	\end{align}
	where $\mathbf{\bar{p}}$ is the average momentum in the given output port, $\Delta \mathbf{p}$ is the momentum separation, which is mapped to the position separation through ballistic expansion as $\Delta \mathbf{p} \rightarrow m\Delta \mathbf{x}/t$ where $t$ is total time allowed for expansion. $\mathbf{x}$ is the position in space, and $\Delta \mathbf{x}$ is the spatial mismatch at the final beamsplitter. This spatial mismatch results in the first term providing a phase offset to the total interferometer phase. For an asymmetric Mach-Zehnder interferometer with a temporal asymmetry, the spatial separation is given by 
	\begin{align}
		\Delta x = v_r \delta T = \frac{2 n\hbar k \Delta T}{m},
	\end{align}
	where $v_r$ is the recoil velocity.

\begin{figure}
		\centering\includegraphics[width=\columnwidth]{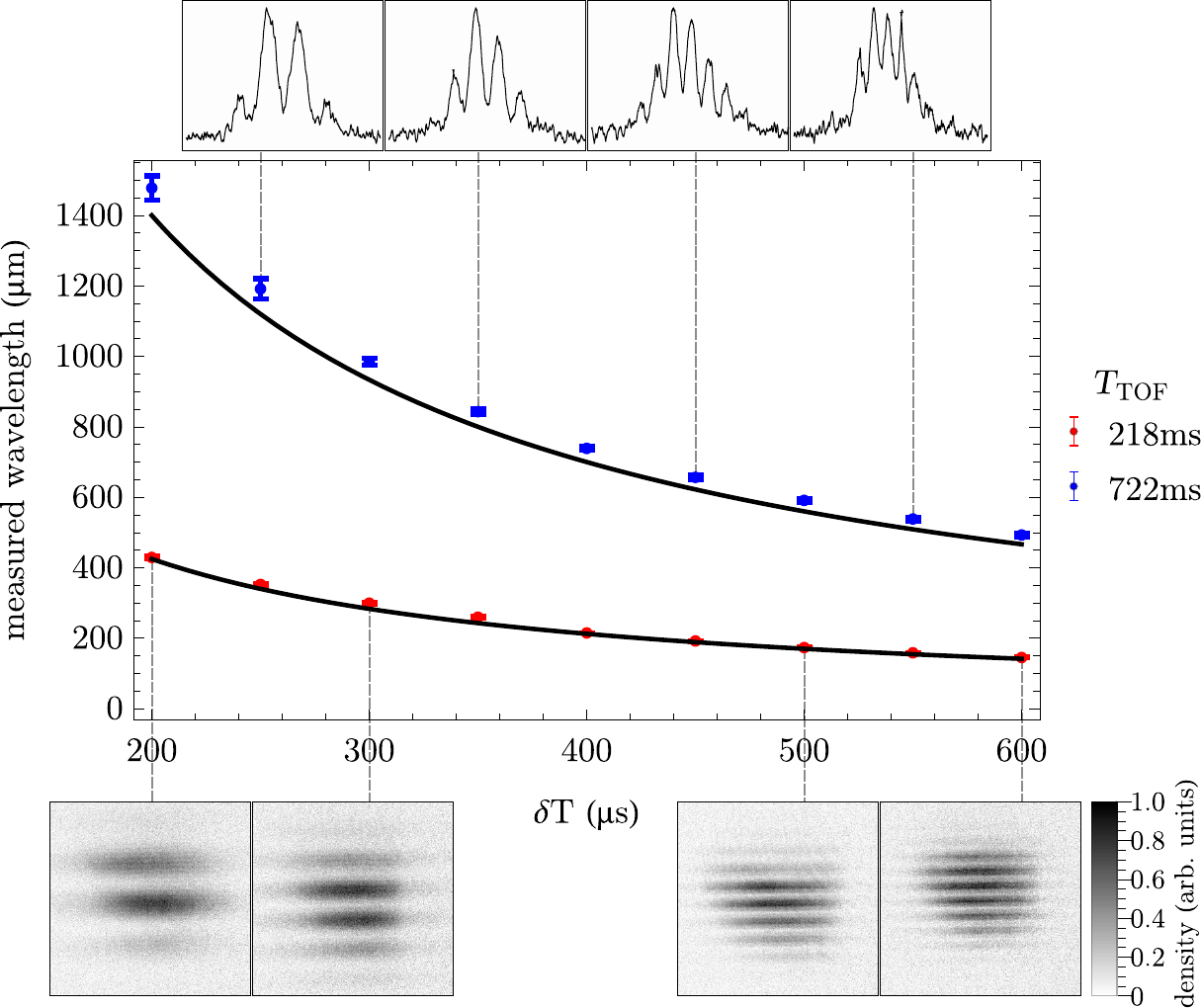}
		\caption{Scaling of the spatial fringe wavelength with the magnitude of the temporal asymmetry and the total time of flight. For $218\unit{ms}$ and $722\unit{ms}$ time of flight, the wavelength of the fringes was measured for a number of values of the temporal asymmetry $\delta T$. The $218\unit{ms}$ data was obtained using absorption imaging, with raw single-run images shown below the plot, while the $722\unit{ms}$ data was obtained using FMI with single run examples shown above the plot. The FMI data has a Savitzky–Golay filter applied. Each data point represents the mean measured wavelength for $100$ runs, with the error bars indicating one standard deviation from the mean. The solid black line shows the analytic expression for the wavelength according to \refeq{fringeWavelength}, dependent on both the temporal asymmetry and the time of flight from release, providing good agreement to the theoretical expression.  }
		\label{fig:kScalingBoth}
	\end{figure}

	The second term in \refeq{separationPhase} results from a spread of momentum across the cloud that, along with the time asymmetry, results in a spatially dependent phase written on at the final beamsplitter. In this case the output states display a sinusoidal modulation on top of the density envelope with a phase that is linked to the inertial acceleration. The wavenumber of this spatial fringe is given by $\Delta p /\hbar$. Converting to a wavelength,
	\begin{align}
		\lambda_\text{fringe} = \frac{2\pi}{k_\text{fringe}} = \frac{2\pi \hbar}{\Delta p}.
	\end{align}
	Here, $\Delta p$ may be mapped to the known quantity $\Delta x$ through ballistic expansion. Each velocity class in the cloud expands at $x/t$, thus mapping $\Delta p \rightarrow m\Delta x/t$ where $t$ is total time allowed for expansion. This results in a spatial fringe wavelength given by
	\begin{align}
		\lambda_\text{fringe} = \frac{\pi T_\text{TOF}}{nk \delta T}
		\label{eq:fringeWavelength}
	\end{align}
	where $T_\text{TOF} = T_0+T_1+T_2+T_\text{sep}$ is the total time of flight after release of the cloud from trap given the timing indicated in \reffig{diagram}. This scaling of the fringe wavelength is shown in \reffig{kScalingBoth} where measured wavelengths are shown for two different expansion times. Example fringes are shown for the $218\unit{ms}$ and $722\unit{ms}$ expansion, along with the curve given by \refeq{fringeWavelength}. A slight deviation exists for the long expansion, likely due to a slightly different expansion rate caused by the Stern-Gerlach pulse used to separate the magnetic substates.

\begin{figure}
		\centering\includegraphics[width=\columnwidth]{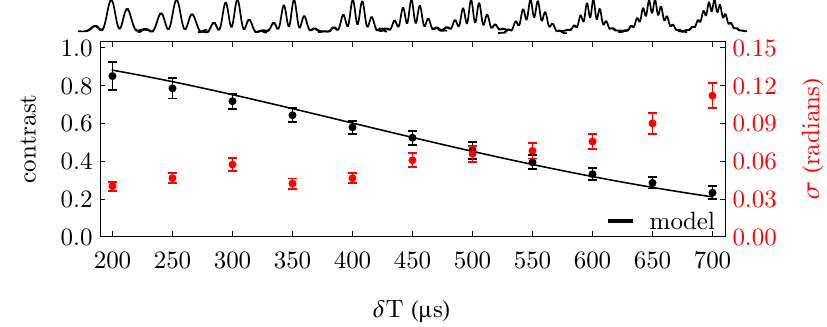}
		\caption{Comparison of the short term phase noise for varying spatial frequencies, obtained by changing the value of the asymmetry, $\delta T$ on a $1\unit{ms}$ interferometer observed using FMI. The phase noise, given by the Allan deviation at a $1$ run averaging time, tends to increase with increasing asymmetry indicating that lower values provide a more optimal operation of the interferometer. The contrast, indicated by the black points, is also seen to decrease as the asymmetry increases, limiting the extent to which the asymmetry can be increased. These features are illustrated by example fringes shown with their associated $\delta T$ on the top horizontal axis. The black line indicates the a model that includes a $5\degree$ rotation of the imaging light relative to the Bragg lattice, suggesting a possible cause of the diminishing contrast. Longer wavelengths are restricted by the size of the cloud limiting the ability to extract a phase value.}
		\label{fig:kspacingComparison}
	\end{figure}

	A decrease in contrast is also observed as the asymmetry increases. This effect may result from a number of sources, both technical and fundamental. In this regime however, the likely candidate for the decrease in contrast is due to the imaging method. Since both techniques are absorptive and provide only an integrated density measurement, if the imaging light is not exactly perpendicular to the fringe pattern the fringes become washed out. This effect is sensitive to both the size of the cloud and the fringe wavelength. A short $T=1\unit{ms}$ interferometer was used to eliminate other sources of noise and directly address the drop in contrast. The FMI system was used to observe the fringes so as to provide a baseline for later phase noise analysis with larger interferometer times. A simple model is shown in \reffig{kspacingComparison} assuming the spatial fringe signal is integrated along one direction with the light at an angle $\theta$ to the fringe pattern. A $5\unit{degree}$ angle would result in the observed decrease in contrast. The contrast is also expected to decrease once the separation of the clouds becomes large compared to the coherence length of the source. The large spatio-temporal coherence of a BEC allows for both larger values of the asymmetry than a thermal cloud in addition to larger interferometer times in the asymmetric setup. The reduction in spatial fringe contrast is shown in the raw signals shown inset in \reffig{kScalingBoth} for both absorption (below) and FMI (above). For this particular setup, the decrease in contrast limits the practical range of $\delta T$ typically to below $1\unit{ms}$, or a spatial separation at the final beamsplitter of ${\sim10}\unit{\upmu m}$. 

	Since the output states cannot be imaged independently, the conventional approach to extracting the phase signal from the spatial fringe measurement involves allowing time for the two output ports to separate before imaging, as shown at the bottom of \reffig{diagram}. The presented technique removes the requirement for state separation, reducing the downtime while simultaneously increasing the possible interferometer duration, directly improving the SWaP capabilities of the device. In order to achieve this, imaging is performed almost immediately after the final beamsplitter of the asymmetric scheme, allowing only enough time for the sinusoidal modulation on each cloud to overlap constructively. That is, a separation time that results in the clouds moving a quarter of a wavelength apart optimizes the overlap of the fringes. This optimal separation time is given by
	\begin{align}
		t_{\pi/2}= \frac{\lambda/4}{v_r} = \frac{1}{8}\frac{m\pi T_\text{TOF}}{n k^2\hbar \delta T}
		\label{eq:overlappedWaitTime}
	\end{align}
	where $v_r = 2n\hbar k/m$ is the recoil velocity. For the $218\unit{ms}$ time of flight, this corresponds to ${\sim}6\unit{ms}$ separation time, and for the $722\unit{ms}$ time of flight this corresponds to ${\sim}15\unit{ms}$ separation time. In contrast, the full separation of the two output ports requires ${\sim100}\unit{ms}$ and ${\sim250}\unit{ms}$ respectively, demonstrating the potential of the scheme for reducing total required freefall time. This would allow devices based on this technique to be significantly reduced in size while maintaining the same interferometer time. By overlapping the spatial fringes the requirement of full separation is relaxed and the phase can be extracted with an increased signal-to-noise. For this technique to be useful for precision measurement applications, the phase noise of the asymmetric interferometer must be at least comparable to the symmetric version and further, the overlapped asymmetric interferometer must not show significant increase in phase noise. 

% \section{Results}\label{sec:results}

    The phase noise for both the symmetric and asymmetric Mach-Zehnder interferometers is calculated using an Allan deviation. This provides a means of quantifying the stability of a signal at various time scales and is defined as the square root of the Allan variance,
    \begin{align}
    	\sigma_y^2\left(\tau\right) = \frac{1}{2\left(M-1\right)}\sum_{i=1}^{M-1}\left(y_{i+1}-y_i\right)^2
    \end{align}
    for a set of $M$ mean data points, $y_i$, obtained at average time $\tau$. For this analysis, the data points correspond to the measured phase and the averaging time is given in units of runs, which correspond to the $11.4\unit{s}$ duty cycle of the experiment. The Allan deviation is a standard tool for assessing the temporal characteristics of noise in precision measurements. The analysis presented here evaluates the short term phase stability of the interferometer schemes, with long term stability to be evaluated in a future publication once complex systematics have been investigated.  

	\begin{figure}
		\centering\includegraphics[width=\columnwidth]{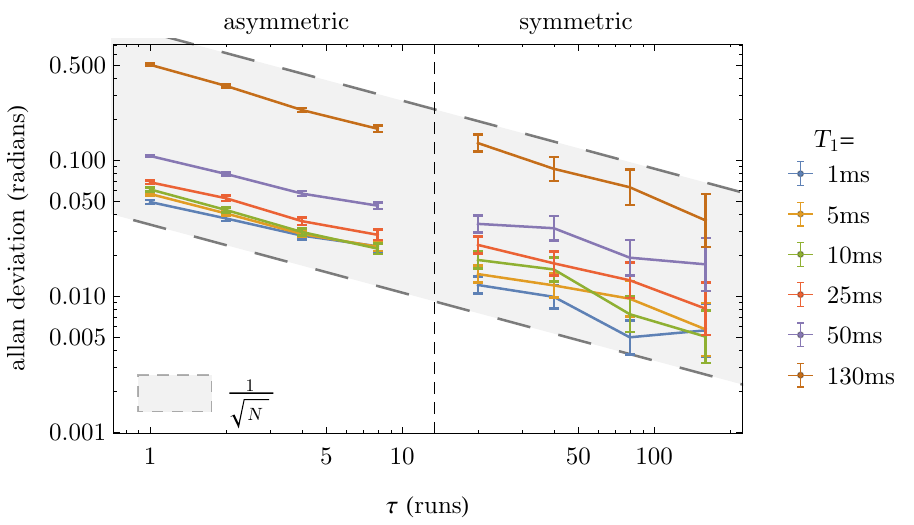}
		\caption{Comparison of the phase noise of the symmetric and asymmetric Mach-Zehnder interferometer as calculated through the Allan deviation. The phase for the symmetric case is obtained by scanning the laser phase then fitting a sinusoid to $20$ points. The phase in the asymmetric case is obtained each run by fitting the sinusoidal modulation. The trend for various interferometer times, $\mbox{T=\{1,\,5,\,10,\,25,\,50,\,130\}}$ms, is seen to be the same between the two methods, with both integrating down according to a $1/\sqrt{\tau}$ trend, as would be expected for white noise. The extrapolated $(\text{Hz})^{-1/2}$ phase noise matches between the two methods. The $T=130\unit{ms}$ represents the phase from an overlapped spatial fringe. }
		\label{fig:bigTScaling}
	\end{figure}

    The phase from the symmetric Mach-Zehnder interferometer is obtained in the usual manner. That is, the signal is boxed around the two states and integrated with the relative population between the two states calculated. This is done for the entire dataset, before binning such that a full fringe can be fitted with a sinusoid of the form 
    \begin{align}
    	f(x)=\mathcal{V} \sin\left(k x+\phi\right)+C
    	\label{eq:symmetricFit}
    \end{align}
    where $k$ represents the frequency of the scan of laser phase, $\mathcal{V}$ represents the visibility, $C$ represents some constant offset and $\phi$ provides the relevant phase value. The resulting phase signal can then be determined and the phase stability calculated. In practice, an interferometer of this type is run mid-fringe, with a new fringe scanned whenever necessary. In this case $20$ runs were required to extract a phase, limiting the minimum averaging time shown in the Allan deviation of \reffig{bigTScaling}. 

    The phase from the spatial fringe signal is obtained by fitting the signal with a function incorporating a Gaussian envelope with a sinusoidal modulation of the form 
    \begin{align}
    	f(x)=A \exp\left[-\frac{\left(x-x_0\right)^2}{2\sigma_x^2}\right]\left[1- B \sin{(k x - \phi)}\right]+C
    \end{align}
    where $y=\{A,\,x_0,\,\sigma_x,\,B,\,k,\,\phi,\,C\}$ are free parameters. Importantly, $k$ represents the spatial frequency, $B$ is the contrast of the spatial fringe and $\phi$ is the phase. An initial fit is performed with the median value of the spatial frequency, $k$, subsequently input to a second round of fitting where this value is held constant. The phase is given by $\phi$ of this final fit, and is taken relative to some arbitrary point in space. This choice of phase reference means that the measurement is now susceptible to vibrations of the camera relative to the Bragg lattice, but the mapping of the lattice frequency to the spatial fringe frequency means that the perturbations are relatively small and are not observed to be a limiting factor in the measurement.

    The Allan deviation of the phase was calculated for various values of the spatial frequency over a $100$ run acquisition, and is shown in \reffig{kspacingComparison} demonstrating the short term phase noise. This indicates that, although higher spatial frequencies would be expected to provide a better determination of phase, the contrast of the fringes decreases correspondingly and results in an increase in phase noise. In addition, a higher spatial frequency increases the contribution of phase noise due to classical effects such as vibrations of the camera shot-to-shot. In general, lower spatial frequencies tend to have better phase stability, though this is limited when the fringe spacing becomes comparable to the size of the cloud. Relatively good phase stability coupled with a favorable ratio of fringe spacing to cloud size is seen for the ${\sim350}\unit{\upmu s}$ asymmetry, with this value being used for the remaining analysis. 

	\begin{figure}
		\centering\includegraphics[width=\columnwidth]{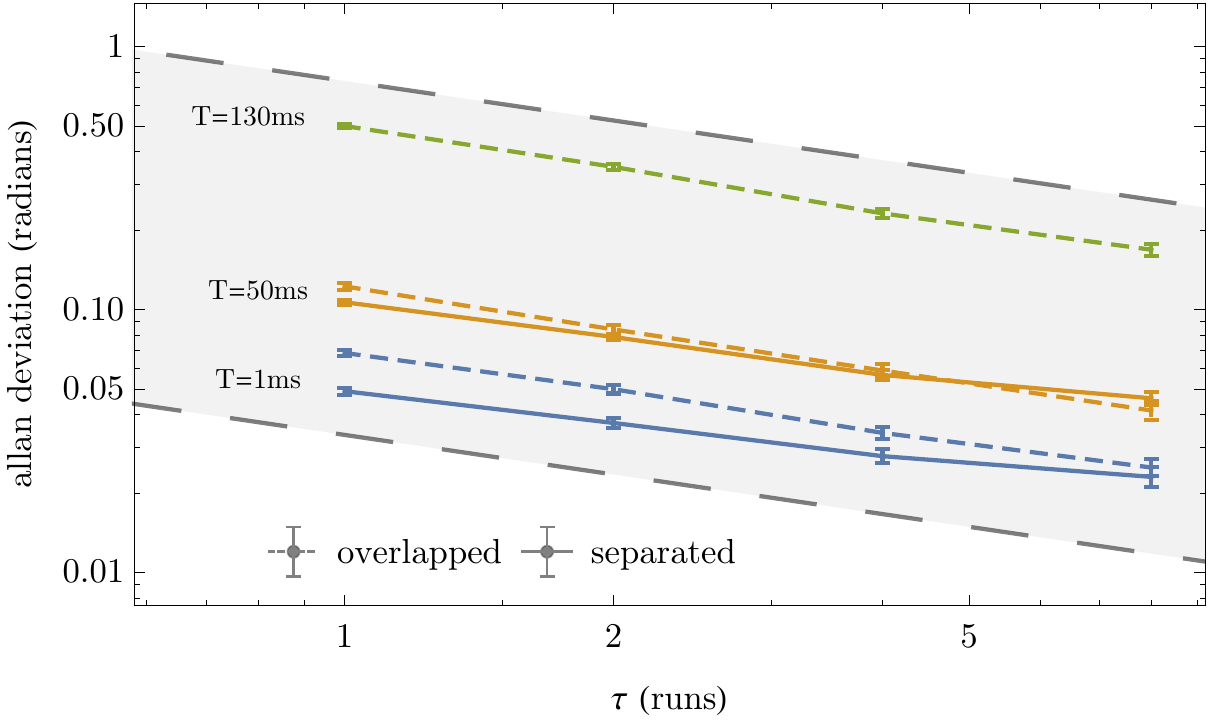}
		\caption{Comparison of the phase noise for an overlapped and non-overlapped asymmetric Mach-Zehnder interferometer for $T=\{1,\,50\}$ms in addition to an overlapped $T=130\unit{ms}$ asymmetric interferometer. The short term phase noise is equivalent between the two methods, suggesting that the overlapped interferometer could be used to mitigate downtime without significantly limiting the phase noise. }
		\label{fig:overlappedComparison}
	\end{figure}
	
    Before investigating the phase noise of the overlapped asymmetric scheme, it is important to verify that the separated asymmetric scheme does not add undesirable phase noise over the conventional symmetric version. In order to compare these two schemes, $1000$ runs were performed for each of of the separated schemes for $T=\{1,\,5,\,10,\,25,\,50,\,130\}\unit{ms}$. The laser phase was scanned in both cases, with the scan in the symmetric version producing the sinusoidal signal described in \refeq{symmetricFit}, enabling the phase to be extracted. The laser phase in the asymmetric case manifests as a linearly increasing phase on the spatial fringe that can subtracted from the signal. The Allan deviation of the phase was determined for both schemes and is shown in \reffig{bigTScaling}. The asymmetric scheme shows the same trend in phase noise with averaging time, with both integrating down following a $1/\sqrt{\tau}$ trend. Furthermore, the extrapolated $(\text{Hz})^{-1/2}$ phase noise for the symmetric scheme matches that of the asymmetric scheme. Whereas the symmetric scheme requires multiple runs to generated a fringe for phase estimation, the asymmetric case allows for single shot phase estimation and requires significantly lower sampling rates without compromising phase noise.

	In order to be useful for precision sensing applications, the phase noise of the overlapped asymmetric interferometer scheme must be competitive with the separated version. In order to test this, $1000$ measurements of phase were taken using the overlapped and separated spatial fringe method. The overlap was achieved by measuring the density signal $19\unit{ms}$ after the final beamsplitter pulse, whereby the fringes from the two output ports added constructively. This required wait time may be calculated given the expected fringe spacing and the momentum difference of the two ports as given by \refeq{overlappedWaitTime}. In practice, this is empirically optimized by monitoring the contrast in the resultant single spatial fringe. The Allan deviation for the overlapped output for $T=\{1,\,50,\,130\}\unit{ms}$ is shown in \reffig{overlappedComparison} along with the Allan deviation of the separated fringes for $T=\{1,\,50\}\unit{ms}$. No significant increase in short-term phase noise is seen to occur due to the overlapping of the fringes, and the $1/\sqrt{\tau}$ scaling remains even for large interferometer times. For this given setup, the overlapped spatial fringe method allows for an increase in interferometer time from a maximum of $T=130\unit{ms}$ limited by separation time, to $T=330\unit{ms}$, significantly improving the potential sensitivity of the measurement. Conversely, this would allow a substantially smaller device while maintaining the same sensitivity, a key requirement for improving SWaP and transitioning such devices from the lab and into the field.

% \section{Conclusion}
    A method for extracting a phase signal from a Bragg based atom interferometer was presented whereby the requirement for state separation is relaxed. This method involves allowing only enough time for the spatial fringes to overlap constructively, greatly reducing the amount of time dedicated to waiting for the interferometer states to separate. This technique allows larger interferometers time for a given sized vacuum system, or equally, smaller vacuum systems required for a given acceleration sensitivity. The phase noise between the asymmetric Mach-Zehnder atom interferometer and the symmetric version is compared and shown to be comparable in the short term limit. An analysis of the phase noise scaling with the magnitude of the asymmetry is presented and shown to be optimum for lower spatial frequencies, limited only by the size of the clouds. Finally, a comparison of the phase noise of large timescale interferometers is compared for overlapped and non-overlapped asymmetric interferometers, showing no increase in short term phase noise. Further work is required to assess long term stability and the susceptibility to drifts arising from the measurement method. The short term noise suggests that an overlapped asymmetric Bragg based Mach-Zehnder atom interferometer could provide a pathway to improving the size, weight and power requirements of portable precision acceleration sensors with the added benefit of single shot phase estimation. In addition, the ability to utilize other, magnetically sensitive substates for simultaneous measurements of magnetic field gradient, and the possibility of extracting a rotation signal from the spatial fringe orientation paves the way for an all-in-one device.

\begin{acknowledgments}
This research is supported by the Commonwealth of Australia as represented by the Defence Science and Technology Group of the Department of Defence. The authors would like to thank Stuart Szigeti for productive and insightful discussions.
\end{acknowledgments}

\bibliographystyle{apsrev4-1}
\bibliography{spatialfringes}

\end{document}